\documentclass[prl,10pt,twocolumn]{revtex4}%
\usepackage{amsfonts}
\usepackage{amsmath}
\usepackage{amssymb}
\usepackage{graphicx}
\usepackage{hyperref}%
\providecommand{\U}[1]{\protect\rule{.1in}{.1in}}

\begin{document}
\title{Quantum effects of Aharonov-Bohm type and noncommutative quantum mechanics}
\author{Miguel E. Rodriguez R.}
\affiliation{Mechatronics Engineering Department, Faculty of Engineering of Applied
Science, Technical University North, Ibarra, 100150-Ecuador}

\pacs{03.65.-w, 2.40.Gh, 03.65.Vf.}
\keywords{Aharonov-Bohm effect, quantum mechanics, noncommutative space--time.}
\begin{abstract}
Quantum mechanics in noncommutative space modifies the standard result of the
Aharonov-Bohm effect for electrons and other recent quantum effects. Here we
obtain the phase in noncommutative space for the Spavieri effect, a generalization of Aharonov-bohm effect which
involving a coherent superposition of particles with opposite charges moving along single open interferometric path. By means of the experimental considerations a
limit $\sqrt{\theta}\simeq\left(  0,13T\text{eV}\right)  ^{-1}$ is
achieved, improving by 10 orders of magnitude the derived by Chaichian
\textit{et. al.} for the Aharonov-Bohm effect. It is also shown that the
noncommutative phases of the Aharonov-Casher and He-McKellar-Willkens effects
are nullified in the current experimental tests.

\end{abstract}
\maketitle

\section{Introduction}

Recently there has been a growing interest in studying quantum mechanics in
noncommutative (NC) space  \cite{Ma2017120}\cite{Ma2016306}\cite{Benchikha2017}%
\cite{Bertolami-2015}\cite{Kovacik-2017}\cite{Ababekri:2016ois}. Because
quantum nature experiments are measured with high precision, these are
feasible scenarios for setting limits on the experimental manifestation of NC space. The Aharonov-Bohm (AB) effect, in which two coherent beams of charged
particles encircle an infinite solenoid \cite{AB}, has been studied by
Chaichian et. al. \cite{Chaichian2002149}, and Li and Dulat \cite{Li2006825}
in the NC space. The expression of the obtained phase includes an additional
term dependent on the NC space parameter, $\theta$ (measured in units of
$(length)^{2}$ ). The limit on $\theta$ found in the AB effect is of the order
of $\sqrt{\theta}\leqslant$ $10^{6}$GeV$^{-1}$ which corresponds to a
relatively large scale of 1\AA \ \cite{Chaichian2002149}. This same approach
was extended to the Aharonov-Casher (AC) \cite{AC} effect by Li and Wang
\cite{Li20071007}, and Mirza and Zarei \cite{Mirza2004583}, in this effect two
coherent beams of neutral particles encircle an infinite charged wire.
Considering the reported experimental error of the AC effect ($\sim25\%$)
\cite{CIMINO-PhysRevLett.63.380} a limit $\sqrt{\theta}\leqslant10^{7}%
$GeV$^{-1}$ is obtained \cite{Mirza2004583} \cite{Li20071007}. The
He-Mckellar-Wilkens (HMW) \cite{He-Mackellar-PhysRevA} \cite{Wilkens} effect,
in which neutral particles with electric dipole moment interact with an
magnetic field, has been studied in the NC context by Wang and Li
\cite{Wang20072197} \cite{Wang20075}, and by Dayi \cite{Dayi2009} and in the
context of the Anandan phase \cite{Anandan-PhysRevLett.85.1354} by Passos
\cite{Pasos-PhysRevA.76.012113}. There is no experimental report on the
parameter limit $\theta$ in quantum effects for electric dipoles. We consider
here a new effect of AB type, proposed by Spavieri in \cite{Spavieri-E-P}. In
this effect, two beams of particles with charges, $+q$ and $-q$, move along a
single side of an infinite solenoid, even though the beams do not enclose the
solenoid (as in the ordinary AB effect). The advantage of this effect, called
here \textquotedblleft the S effect\textquotedblright, is that the size of the
solenoid has no limit, so that it can be considered to be very large, such as
a cyclotron. The S effect has been studied by Spavieri and Rodriguez
\cite{spavieri:052113} in the context of massive electrodynamics (or photon
mass). Under certain experimental considerations proposed and discussed in
\cite{spavieri:052113}, Spavieri and Rodriguez envisage a limit on the mass of
the photon of $m_{\gamma}\sim10^{-51}$g, which is the best limit obtainable
for the photon mass by means of a laboratory experiment with a quantum
approach. Consequently, due to the success of the S effect in the photon mass
scenario, we derive here the phase of the S effect in the context of the NC quantum mechanics
 as an application of the phase found by Chaichian et. al.
\cite{Chaichian2002149}. Keeping the experimental proposal of
\ \cite{spavieri:052113} we get a new limit on $\theta$ in the context of the
quantum effects of the AB type. In addition, recent advances in atomic
interferometry have allowed  obtaining measurements of the HWM phase
\cite{Lepoutre2013} which allows  exploring experimentally the manifestation
of the space  NC in the HWM effect by means of the phases found in
\cite{Pasos-PhysRevA.76.012113} for these effects. This same analysis may be
extended to the experimental configuration proposed by Sangster et. al.
\cite{Sangster-PhysRevLett.71.3641} for the AC effect where the particles do
not enclose the charged wire.
\vspace{-0.5cm}

\section{NC quantum mechanics }

In NC quantum mechanics, the commutation relationships of the position
operators satisfy the relation, $\left[  \hat{x}_{i},\hat{x}_{j}\right]
=$i$\theta_{ij}$, where $\left\{  \theta_{ij}\right\}  $ is a fully
antisymmetric real matrix representing the property noncommutativity of space
and $\hat{x}_{i}$ represents the coordinate operator ($\hat{p}_{i}$ is the
corresponding moment operator) in the space NC. In this scenario the product
of two functions is replaced by the product Moyal-Weyl (or star
\textquotedblleft$\ast$\textquotedblright\ ) \cite{MANKO2007522}, so the
ordinary Schr\"{o}ndiger equation, $H\psi=E\psi$, is written as:
\begin{equation}
H\left(  \hat{x}_{i},\hat{p}_{i}\right)  \ast\psi=E\psi
\label{Schrodinger-NCCM}%
\end{equation}

Now, the star product between two functions in an NC plane $(i,j=1,2)$ is
defined by the following expression:%

\begin{align}
\left(  f\ast g\right)  \left(  x\right)   &  =e^{\frac{\text{i}}{2}%
\theta_{ij}\partial_{x_{i}}\partial_{x_{j}}}f\left(  x_{i}\right)  g\left(
x_{j}\right) \label{Moyal-Weyl}\\
&  =f\left(  x\right)  g\left(  x\right)  +\left.  \frac{\text{i}}{2}%
\theta_{ij}\partial_{i}f\partial_{j}g\right\vert _{x_{i}=x_{j}}+O\left(
\theta^{2}\right)  ,\nonumber
\end{align}

where $f\left(  x\right)  $ and $g\left(  x\right)  $ are two arbitrary
functions. Usually, the NC operators are expressed by means of the formulation
of the Bopp shift \cite{Bopp-Shift-DULAT-2006}\ (equivalent to
(\ref{Moyal-Weyl})). This formalism maps the NC problem in the usual
commutative space using new NC variables defined in terms of the commutative
variables. That is to say,%

\begin{equation}
\hat{x}_{i}=x_{i}-\frac{1}{2\hbar}\theta_{ij}p_{j}\text{, \ \ }i,j=1,2,
\label{coordenadas-NCMC}%
\end{equation}

where the variables $x_{i}$ and $p_{i}$ satisfy the usual canonical
commutation relations, $\left[  x_{i},x_{j}\right]  =0$, $\left[  x_{i}%
,p_{j}\right]  =i\hbar\delta_{ij}$ and $\left[  p_{i},p_{j}\right]  =0.$ With
these considerations the Hamiltonian undergoes a coordinate transformation,
$H\left(  \hat{x}_{i},\hat{p}_{i}\right)  =H\left(  x_{i}-\frac{1}{2\hbar
}\theta_{ij}p_{j},p_{i}\right)  $. Note that $\theta_{ij}\ll1$, so that the
effects of the NC space can always be treated as a disturbance. If we consider
a particle of mass $m$ and charge $q$ in the presence of a magnetic field (or
potential vector $A_{i}$), then the Hamiltonian in the space NC, $H(
\hat{x}_{i},\hat{p}_{i},\hat{A}_{i})  $ undergoes a Bopp shift in both
$\hat{x}_{i}$ and $\hat{A}_{i}$. Therefore, in the NC space and with a
magnetic field the equation of Schr\"{o}dinger takes the following form:%

\begin{equation}
\frac{\hbar^{2}}{2m}\left(  p_{i}-qA_{i}-\frac{1}{2}q\theta_{lj}p_{l}%
\partial_{j}A_{i}\right)  ^{2}\psi=E\psi, \label{schrodinger-NC-final-1}%
\end{equation}
whose solution is:%

\begin{equation}
\psi=\psi_{0}\exp\left[  \text{i}\frac{q}{\hbar}\int_{x_{0}}^{x}\left(
A_{i}+\frac{1}{2}\theta_{lj}p_{l}\partial_{j}A_{i}\right)  dx_{i}\right]  ,
\label{fase-AB}%
\end{equation}

where $\psi_{0}$ is the solution of
(\ref{schrodinger-NC-final-1}) when $A_{i}=0$.
\vspace{-0.5cm}

\section{Phase  of Effect S in NC quantum mechanics and limit over $\theta$}

In \cite{Spavieri-E-P} Spavieri has pointed out that the amount observable in
the AB effect is actually the phase difference%

\begin{equation}
\Delta\varphi=\frac{e}{\hbar}\left[\int \mathbf{A}\cdot d\mathbf{l-}%
\int\mathbf{A}_{0}\cdot d\mathbf{l}\right]  , \label{phase-shift-S}%
\end{equation}

where the integral can be taken over an open path integral. For the usual
closed path $C$ encircling the solenoid and limiting the surface $S$, the
observable quantity is the phase-shift variation, \ $\Delta\phi\propto
\oint_{C}\mathbf{A}\cdot d\mathbf{l-}\oint_{C}\mathbf{A}_{0}\cdot
d\mathbf{l=}\oint_{C}\mathbf{B}\cdot d\mathbf{S-}\oint_{C}\mathbf{B}_{0}\cdot
d\mathbf{S.}$ In fact, in interferometric experiments involving the AB and AC
effects \cite{CHAMBERS-PhysRevLett.5.3} \cite{Sangster-PhysRevLett.71.3641}
the direct measurement of the phase $\varphi\propto\int\mathbf{A}\cdot
d\mathbf{l}$ or phase shift $\phi\propto\oint\mathbf{A}\cdot d\mathbf{l}$~is
impossible in principle without the comparison of the actual interference
pattern, due to $\mathbf{A}$, with an interference reference pattern, due to  $\mathbf{A}_{0}$. Thus, $\varphi$ o $\phi$ are
not observable, but the variations $\Delta\varphi$ and $\Delta\phi$ are both
gauge-invariant observable quantities \cite{Spavieri-E-P}. Therefore, with these considerations  introduced by Spavieri \cite{Spavieri-E-P}, it is possible consider a new effect of AB type without particles encircling of solenoid. In this case the particles must has opposites charges,  $\pm e$ , moving along one side of solenoid, i. e., along path $b$. Thus, the phase of this new effect called Spavieri effect, S, is:

\begin{equation}
\Delta\varphi_{S}=\frac{e}{\hbar}\left[\int_{b} \mathbf{A}\cdot d\mathbf{l-}%
\int_{b}\mathbf{A}\cdot d\mathbf{l}\right]= \frac{2e}{\hbar}\int_{b}\mathbf{A}\cdot d\mathbf{l}. \label{phase-shift-S1}%
\end{equation}
Now, we interesting in the phase (\ref{phase-shift-S1}) in the context of NC  quantum mechanics.  Substituting the phase (\ref{fase-AB}) in (\ref{phase-shift-S1}), and retaining only term related with the parameter $\theta$, we obtained the correction due to NC space for the effect S, thus:
\begin{equation}
\Delta\varphi_{S}^{NC} =  \frac{e}{\hbar}\int_{b}
\theta_{lj}p_{l}\partial_{j}A_{i}  dx_{i}. \label{phase-shift-S2}%
\end{equation}
 Writing (\ref{phase-shift-S2}) en Cartesian coordinates, we obtained, the phase-shift of effect S in NC space:
\begin{equation}
\Delta\phi_{S}^{NC}=-\frac{em}{4\hbar^{2}}\vec{\theta}\cdot\int\left[  \left(
\mathbf{v}\times\vec{\nabla}A_{i}\right)  -\frac{e}{m}\left(  \mathbf{A}%
\times\vec{\nabla}A_{i}\right)  \right]  dx_{i}. \label{fase-S-NC}%
\end{equation}
where $i=1,2$ are Cartesian components $x$ and $y$, $m$  is the mass of electron and $\mathbf{v}$ is the velocity of particles.
Although the effect for $\pm q$ charged particles is viable
\cite{Spavieri-E-P}, the technology and interferometry for the test of this
effect needs improvements. It is worth recalling that not long ago the
technology and interferometry for beams of particles with opposite magnetic
$\pm\mathbf{m}$ or electric $\pm \mathbf{d} $ dipole moments was likewise
unavailable, but is today a reality \cite{Sangster-PhysRevLett.71.3641}%
\cite{DWF-PhysRevLett.83.2486}. Discussions on this subject may act as a
stimulating catalyst for further studies and technological advances that will
lead to the experimental test of this quantum effect. An important step in
this direction has already been made \cite{Spavieri-E-P} by showing that, at
least in principle and as far as gauge invariance requirements are concerned,
this effect is physically feasible. 

In the experimental setups detecting the traditional AB effect there are
limitations imposed by the suitable type of interferometer related to the
electron wavelength, the corresponding convenient size of the solenoid or
toroid, and the maximum achievable size $\rho$ of the coherent electron beam
encircling the magnetic flux \cite{BD-PhysRevLett.63.2319}. In the analysis
made by Boulware and Deser \cite{BD-PhysRevLett.63.2319} in the context of the limit of mass photon, the radius of the
solenoid is $a=0.1$cm, and $\rho$ is taken to be about 10cm, implying that the
electron beam keeps its state of coherence up to a size $\rho=10^{2}a$, i.e.,
fifty times the solenoid diameter. The advantage of the new approach for the
$\pm q$ beam of particles is that the dimension of the solenoid has no upper
limits and is conditioned only by practical limits of the experimental setup,
while the size of the coherent beam of particles plays no important role. Due
to these advantages of the new approach introduced by Spavieri and its success
in the exploration of the limit of the mass of the photon, it is asked which
limit could be reached for the parameter $\theta$ of the NC quantum mechanics. To answer this question, we consider that the electrons move along the straight line $y=y_{0}$ from $x=-x_{0}$ to
$x=x_{0}$ (open path $b$ in (\ref{phase-shift-S2})), with a velocity $\mathbf{v}=v\mathbf{i}$, thus $i=x$ in
(\ref{fase-S-NC}). In addition, as in \cite{Chaichian2002149}, here it is
considered that $\vec{\theta}$ $=\theta\mathbf{z}$. To complete the calculation it is necessary to know the component $A_{x}$ of
the external vector potential to an infinite solenoid,%

\[
A_{x}=-B_{0}\frac{a^{2}}{2}\left(  \frac{y}{x^{2}+y^{2}}\right)  ,
\]

thus, the terms in parentheses of (\ref{fase-S-NC}) are:%
\begin{equation}
\mathbf{v}\times\vec{\nabla}A_{x}=-B_{0}\frac{a^{2}v}{2}\left(  \frac{
x^{2}-y^{2} }{\left(  x^{2}+y^{2}\right)  ^{2}}\right)  \mathbf{z}
\label{termino-A-1}%
\end{equation}

and%

\begin{equation}
\mathbf{A}\times\vec{\nabla}A_{x}=-\frac{1}{4}\frac{B_{0}^{2}a^{4}y}{\left(
x^{2}+y^{2}\right)  ^{2}}\mathbf{z} \label{termino-A-2}%
\end{equation}

where $a$ is the radius of  solenoid and $B_{0}$ is the magnetic field enclosed within of solenoid. Substituting (\ref{termino-A-1}) and (\ref{termino-A-2}) in
(\ref{fase-S-NC}) and performing the integration of $-x_{0}$ to
$-x_{0}$ the correction NC to the phase of the effect S  is obtained:

\begin{equation}
\Delta\phi_{S}^{NC}=\frac{1}{8}\theta\left(  \frac{\Phi}{\Phi_{0}}\right)
^{2}\left\{
\begin{array}
[c]{c}%
  \dfrac{\arctan\left(  \dfrac{x}{y}\right)  }{y^{2}}+\dfrac{x/y}%
{x^{2}+y^{2}} \\ \\
+\dfrac{8\pi}{\lambda_{e}}\dfrac{\Phi_{0}}{\Phi}\dfrac{v}{c}\dfrac{x}{x^{2}+y^{2}}%
\end{array}
\right\}  \label{phase-S-NCQM}%
\end{equation}

where $\Phi=\pi a^{2}B_{0}$ is the magnetic flux enclosed within solenoid, $\Phi_{0}=2,06\times10^{-15}$T$\cdot$m$^{2}$ is the quantum flux
elemental, $\lambda_{e}$ the Compton wavelength of the electron and $c$ is the speed of light. To
estimate a limit on $\theta$ here we consider the same experimental parameters
introduced and discussed in \cite{spavieri:052113} for the study of the mass of the photon
in the context of the effect S, which are: $a=5$m$,$ $x=5a=30$m$,$ $y=8a/5=8m$
y $B_{0}=10T$. With these parameters it can be demonstrated that the order of
magnitude (in units $m^{-2}$) of the terms in square brackets are the following: $\frac
{\arctan\left(  \frac{x}{y}\right)  }{y^{2}}\sim10^{-2}$, $\frac{x/y}%
{x^{2}+y^{2}}\sim10^{-3}$ $\ $y $\frac{8\pi}{\lambda_{e}}\frac{\Phi_{0}}{\Phi
}\frac{v}{c}\frac{x}{x^{2}+y^{2}}\sim3,3\times10^{-15}v$. If the velocity of
electrons is $v=2\times10^{8}$m/\thinspace s as in the Tonomura et. al.
\cite{Tonomura1987639} experiment for the Aharonov-Bohm effect, then the order
of magnitude of the kinetic term is\ $10^{-7}$. This analysis shows that the
kinetic term is up to 5 times smaller than the geometric terms. This contrasts
with the analysis made by Chiachian et. al. \cite{Chaichian2002149} for the AB
effect where the kinetic term is five orders greater than the geometric term.
Consequently, for the estimation of the limit on $\theta$ we consider only the
first term in brackets of expression (\ref{phase-S-NCQM}), i.e. $\Delta
\phi_{S}^{NC}\simeq\frac{1}{8}\theta\left(  \frac{\Phi}{\Phi_{0}}\right)
^{2}\frac{\arctan\left(  \frac{x}{y}\right)  }{y^{2}}$. As the NC
correction is a very small, its effect must be masked within experimental
error, $\epsilon$, so $\Delta
\phi_{S}^{NC}\leqslant \epsilon$.  This same argument is followed in the works related to the estimation
of the mass of the photon \cite{BD-PhysRevLett.63.2319}%
\cite{Neyenhuis-Mass-Photon}\cite{spavieri:052113} \cite{Rodriguez2009373}. According to recent
advances in atomic interferometry \cite{1674-1056-24-5-053702,gustavson-2000}, the
experimental error that can reach in the measurement of the quantum phases is
of the order of $10^{-4}$rad, this can be seen in the  measurement of
the  AC
\cite{Sangster-PhysRevLett.71.3641}, where the phase has been measured with an experimental error of $0.11$mrad$=1.1\times10^{-4}%
$rad, even Zhout et. al. \cite{Zhou-2016}, by means of simulation, provided for the measurement of the AC phase with a relative error of $10^{-5}$rad. Consequently, in this work, to be conservative, it is considered that
 $\epsilon=10^{-4}$rad. Therefore the estimated limit on
$\theta$ in the context of the effect S is,%

\[
\sqrt{\theta}\leqslant\left[  \frac{1}{8y}\left(  \frac{\Phi}%
{\Phi_{0}}\right)\sqrt{\frac{\arctan\left(  \frac{x}{y}\right)  }{\epsilon
}} \, \right]  ^{-1}\simeq\left[  0,13\times\text{TeV}\right]  ^{-1}%
\]

which is 10 orders of magnitude smaller than the value of Chiachian et. al.
\cite{Chaichian2002149} for the  effect AB.
\vspace{-0.5cm}
\section{Quantum effect for electric and magnetic dipoles in NC quantum mechanics}

The phase to magnetic dipoles, $\mathbf{m}$ (effect AC) and for electric dipoles, $\mathbf{d}$, (effect
HMW) in noncommutative space also has been calculated by Passos et al.
\cite{Pasos-PhysRevA.76.012113}. The expressions are as follows:%

\begin{align}
\phi_{AC}  &  =\text{i}\oint\left(  \mathbf{m}\times\mathbf{E}\right)
\cdot d\mathbf{r}+\frac{\text{i}}{2}m\oint\vec{\theta}\mathbf{\cdot}\left[
\mathbf{v\times\nabla\cdot}\left(  \mathbf{m}\times\mathbf{E}\right)
\right]  \cdot d\mathbf{r}\nonumber\\
&  \mathbf{-}\frac{\text{i}}{2}m\oint\vec{\theta}\mathbf{\cdot}\left[  \left(
\mathbf{m}\times\mathbf{E}\right)  \mathbf{\times\nabla\cdot}\left(
\mathbf{m}\times\mathbf{E}\right)  \right]  \cdot d\mathbf{r}
\label{AC-passos}%
\end{align}

\begin{align}
\phi_{HMW}  &  =-\text{i}\oint\left(  \mathbf{d}\times\mathbf{B}\right)  \cdot
d\mathbf{r-}\frac{\text{i}}{2}m\oint\vec{\theta}\mathbf{\cdot}\left[
\mathbf{v\times\nabla\cdot}\left(  \mathbf{d}\times\mathbf{B}\right)  \right]
\cdot d\mathbf{r}\nonumber\\
&  \mathbf{+}\frac{\text{i}}{2}m\oint\vec{\theta}\mathbf{\cdot}\left[  \left(
\mathbf{d}\times\mathbf{B}\right)  \mathbf{\times\nabla\cdot}\left(
\mathbf{d}\times\mathbf{B}\right)  \right]  \cdot d\mathbf{r},
\label{hwm-passos}
\end{align}
where $m$ is the mass of electric or magnetic dipoles. In the experimental setup proposed by Sangster et. al.
\cite{Sangster-PhysRevLett.71.3641}, magnetic dipoles with opposing dipole
moments are moving on the same interferometric path. With this configuration
of magnetic moments the beams do not need to enclose a charged wire (as in the
ordinary AC effect), but are moving in the presence of a homogeneous electric
field produced by a capacitor of parallel plates. In the configuration of
Sansgter \cite{Sangster-PhysRevLett.71.3641} the magnetic dipole moments,
$\mathbf{m}$, are perpendicular to the electric field, $\mathbf{E}$,
thus the terms $\mathbf{\nabla\cdot}\left(  \mathbf{m}\times
\mathbf{E}\right)  $ in (\ref{AC-passos}) vanish and the effects of
the NC space can not be observed in this configuration. In the same sense, in a recent
experiment carried out to observe the HMW phase \cite{Lepoutre2013}, the
electric dipole moments, $\mathbf{d}$, of the beams are perpendicular to the
magnetic field, $\mathbf{B}$, involved in the effect. Thus, the terms
$\mathbf{\nabla\cdot}\left(  \mathbf{d}\times\mathbf{B}\right)  $ in the
expression (\ref{hwm-passos}) vanish and the NC effect, as a function of the
expression (\ref{hwm-passos}) derived by Passos et. al. (\ref{hwm-passos}), is
not evidentiable in this configuration. Another proposed configuration for
observing the AB effect for electric dipoles is known as Takchuk effect  \cite{Takchuk-PhysRevA.62.052112}. In this configuration, it is consider
two infinite wires with an opposite magnetic polarization dependent on the
length of the wire, that is, $M\left(  z\right)  =-qz$, where $q$ can be
treated as a linear magnetic charge density. If the wires are sufficiently
long, the magnetic vector potential can be written as $\mathbf{A}%
_{T}=z\mathbf{A}_{AB}$, $\mathbf{A}_{AB}$ being the ordinary vector potential
of the AB effect. Thus, the magnetic field is $\mathbf{B}=\nabla
\times\mathbf{A}_{T}=z\left(  \nabla\times\mathbf{A}_{AB}\right)
-\mathbf{A}_{AB}\times\mathbf{z}$. In this effect the beams of magnetic
dipoles move in the middle plane of the wires, that is, $z=0$, with their
polarization, $\mathbf{d}$, parallel to the wire axis. The term of interest in
the NC context according to (\ref{hwm-passos}) is $\mathbf{d}\times\mathbf{B}%
$. Therefore, $\mathbf{d}\times\mathbf{B=}d$ $\mathbf{z}\times\left(  z\left(
\nabla\times\mathbf{A}_{AB}\right)  -\mathbf{A}_{AB}\times\mathbf{z}\right)
$, evaluated at $z=0$, we obtain that $\mathbf{d}\times\mathbf{B=}%
d\mathbf{A}_{AB}$, but $\nabla\cdot\left(  \mathbf{d}\times\mathbf{B}\right)
=d\nabla\cdot\mathbf{A}_{AB}=0$ for the ordinary AB effect (Coulomb's gauge).
This implies that in the scenario presented by Takchuk
\cite{Takchuk-PhysRevA.62.052112}, the NC effect can not be observed.
\vspace{-0.5cm}
\section{Conclusions}

We have derived the AB phase for open path, or S effect,  in the context of NC quantum
mechanics (eq. \ref{phase-S-NCQM}). Considering the experimental  parameters of the S effect to obtained a limit on mass photon, here we obtained an upper limit on the
NC parameter, $\sqrt{\theta}\leqslant\left(  0.13\text{TeV}\right)  ^{-1}$, 10
orders of magnitude smaller than in previous scenarios of the AB type, and three order of magnitude smaller that the limit derived
by Moumni et. al. \cite{Moumni2011151} $\sqrt{\theta}\leqslant\left(
0.16\text{GeV}\right)  ^{-1}$, in the context of the energy lines of the
hydrogen atom, which is also a quantum scenario. It can also be observed that
the kinetic term in our result, which includes the speed and mass of the
particle, is not relevant for $\theta$ calculation since it is several orders
of magnitude smaller than the geometric terms, which is opposite to the result
found Chaichian et. al. \cite{Chaichian2002149}. Also we have shown that the
NC effects are not manifested in the experimental configurations (for to interferometric open path) proposed by
Sangster et. al. \cite{Sangster-PhysRevLett.71.3641} for the AC effect and the
proposal of Lepoutre et. al. \cite{Lepoutre2013}\cite{Lepoutre2012} and
Takchuk \cite{Takchuk-PhysRevA.62.052112} for the HMW effect from the point of
view of the phases derived by Passos et. al. \ \cite{Pasos-PhysRevA.76.012113}
for these purposes. Finally, it is important to mention that these results can be improved in the future due to the development of new atomic interferometers, with more precision and longer interferometric paths \cite{Wang-Jin-2015}.


\begin{thebibliography}{39}
	\expandafter\ifx\csname natexlab\endcsname\relax\def\natexlab#1{#1}\fi
	\expandafter\ifx\csname bibnamefont\endcsname\relax
	\def\bibnamefont#1{#1}\fi
	\expandafter\ifx\csname bibfnamefont\endcsname\relax
	\def\bibfnamefont#1{#1}\fi
	\expandafter\ifx\csname citenamefont\endcsname\relax
	\def\citenamefont#1{#1}\fi
	\expandafter\ifx\csname url\endcsname\relax
	\def\url#1{\texttt{#1}}\fi
	\expandafter\ifx\csname urlprefix\endcsname\relax\def\urlprefix{URL }\fi
	\providecommand{\bibinfo}[2]{#2}
	\providecommand{\eprint}[2][]{\url{#2}}
	
	\bibitem[{\citenamefont{Ma and Wang}(2017)}]{Ma2017120}
	\bibinfo{author}{\bibfnamefont{K.}~\bibnamefont{Ma}} \bibnamefont{and}
	\bibinfo{author}{\bibfnamefont{J.-H.} \bibnamefont{Wang}},
	\bibinfo{journal}{Ann. Phys.} \textbf{\bibinfo{volume}{383}},
	\bibinfo{pages}{120 } (\bibinfo{year}{2017}).
	
	\bibitem[{\citenamefont{Ma et~al.}(2016)\citenamefont{Ma, Wang, and
			Yang}}]{Ma2016306}
	\bibinfo{author}{\bibfnamefont{K.}~\bibnamefont{Ma}},
	\bibinfo{author}{\bibfnamefont{J.-H.} \bibnamefont{Wang}}, \bibnamefont{and}
	\bibinfo{author}{\bibfnamefont{H.-X.} \bibnamefont{Yang}},
	\bibinfo{journal}{Phys. Lett. B} \textbf{\bibinfo{volume}{759}},
	\bibinfo{pages}{306 } (\bibinfo{year}{2016}).
	
	\bibitem[{\citenamefont{Benchikha et~al.}(2017)\citenamefont{Benchikha, Merad,
			and Birkandan}}]{Benchikha2017}
	\bibinfo{author}{\bibfnamefont{A.}~\bibnamefont{Benchikha}},
	\bibinfo{author}{\bibfnamefont{M.}~\bibnamefont{Merad}}, \bibnamefont{and}
	\bibinfo{author}{\bibfnamefont{T.}~\bibnamefont{Birkandan}},
	\bibinfo{journal}{Mod. Phys. Lett.} \textbf{\bibinfo{volume}{A32}},
	\bibinfo{pages}{1750106} (\bibinfo{year}{2017}).
	
	\bibitem[{\citenamefont{Bertolami and Leal}(2015)}]{Bertolami-2015}
	\bibinfo{author}{\bibfnamefont{O.}~\bibnamefont{Bertolami}} \bibnamefont{and}
	\bibinfo{author}{\bibfnamefont{P.}~\bibnamefont{Leal}},
	\bibinfo{journal}{Phys. Lett. B} \textbf{\bibinfo{volume}{750}},
	\bibinfo{pages}{6 } (\bibinfo{year}{2015}).
	
	\bibitem[{\citenamefont{Kov\'a\v{c}ikand and
			Pre\v{s}najder}(2017)}]{Kovacik-2017}
	\bibinfo{author}{\bibfnamefont{S.}~\bibnamefont{Kov\'a\v{c}ikand}}
	\bibnamefont{and}
	\bibinfo{author}{\bibfnamefont{P.}~\bibnamefont{Pre\v{s}najder}},
	\bibinfo{journal}{J. Math. Phys.} \textbf{\bibinfo{volume}{58}},
	\bibinfo{pages}{012101} (\bibinfo{year}{2017}).
	
	\bibitem[{\citenamefont{Ababekri et~al.}(2016)\citenamefont{Ababekri, Abduwali,
			Mamatabdulla, and Reyima}}]{Ababekri:2016ois}
	\bibinfo{author}{\bibfnamefont{M.}~\bibnamefont{Ababekri}},
	\bibinfo{author}{\bibfnamefont{A.}~\bibnamefont{Abduwali}},
	\bibinfo{author}{\bibfnamefont{H.}~\bibnamefont{Mamatabdulla}},
	\bibnamefont{and} \bibinfo{author}{\bibfnamefont{R.}~\bibnamefont{Reyima}},
	\bibinfo{journal}{Front. Phys.} \textbf{\bibinfo{volume}{4}},
	\bibinfo{pages}{1} (\bibinfo{year}{2016}).
	
	\bibitem[{\citenamefont{Aharonov and Bohm}(1959)}]{AB}
	\bibinfo{author}{\bibfnamefont{Y.}~\bibnamefont{Aharonov}} \bibnamefont{and}
	\bibinfo{author}{\bibfnamefont{D.}~\bibnamefont{Bohm}},
	\bibinfo{journal}{Phys. Rev.} \textbf{\bibinfo{volume}{115}},
	\bibinfo{pages}{485} (\bibinfo{year}{1959}).
	
	\bibitem[{\citenamefont{Chaichian et~al.}(2002)\citenamefont{Chaichian,
			Pre\v{s}najder, Sheikh-Jabbari, and Tureanu}}]{Chaichian2002149}
	\bibinfo{author}{\bibfnamefont{M.}~\bibnamefont{Chaichian}},
	\bibinfo{author}{\bibfnamefont{P.}~\bibnamefont{Pre\v{s}najder}},
	\bibinfo{author}{\bibfnamefont{M.}~\bibnamefont{Sheikh-Jabbari}},
	\bibnamefont{and} \bibinfo{author}{\bibfnamefont{A.}~\bibnamefont{Tureanu}},
	\bibinfo{journal}{Phys. Lett. B} \textbf{\bibinfo{volume}{527}},
	\bibinfo{pages}{149} (\bibinfo{year}{2002}).
	
	\bibitem[{\citenamefont{Li and Dulat}(2006)}]{Li2006825}
	\bibinfo{author}{\bibfnamefont{K.}~\bibnamefont{Li}} \bibnamefont{and}
	\bibinfo{author}{\bibfnamefont{S.}~\bibnamefont{Dulat}},
	\bibinfo{journal}{Eur. Phys. J. C} \textbf{\bibinfo{volume}{46}},
	\bibinfo{pages}{825} (\bibinfo{year}{2006}).
	
	\bibitem[{\citenamefont{Aharonov and Casher}(1984)}]{AC}
	\bibinfo{author}{\bibfnamefont{Y.}~\bibnamefont{Aharonov}} \bibnamefont{and}
	\bibinfo{author}{\bibfnamefont{A.}~\bibnamefont{Casher}},
	\bibinfo{journal}{Phys. Rev. Lett.} \textbf{\bibinfo{volume}{53}},
	\bibinfo{pages}{319} (\bibinfo{year}{1984}).
	
	\bibitem[{\citenamefont{Li and Wang}(2007)}]{Li20071007}
	\bibinfo{author}{\bibfnamefont{K.}~\bibnamefont{Li}} \bibnamefont{and}
	\bibinfo{author}{\bibfnamefont{J.}~\bibnamefont{Wang}},
	\bibinfo{journal}{Eur. Phys. J. C} \textbf{\bibinfo{volume}{50}},
	\bibinfo{pages}{1007} (\bibinfo{year}{2007}).
	
	\bibitem[{\citenamefont{Mirza and Zarei}(2004)}]{Mirza2004583}
	\bibinfo{author}{\bibfnamefont{B.}~\bibnamefont{Mirza}} \bibnamefont{and}
	\bibinfo{author}{\bibfnamefont{M.}~\bibnamefont{Zarei}},
	\bibinfo{journal}{Eur. Phys. J. C} \textbf{\bibinfo{volume}{32}},
	\bibinfo{pages}{583} (\bibinfo{year}{2004}).
	
	\bibitem[{\citenamefont{Cimmino et~al.}(1989)\citenamefont{Cimmino, Opat,
			Klein, Kaiser, Werner, Arif, and Clothier}}]{CIMINO-PhysRevLett.63.380}
	\bibinfo{author}{\bibfnamefont{A.}~\bibnamefont{Cimmino}},
	\bibinfo{author}{\bibfnamefont{G.~I.} \bibnamefont{Opat}},
	\bibinfo{author}{\bibfnamefont{A.~G.} \bibnamefont{Klein}},
	\bibinfo{author}{\bibfnamefont{H.}~\bibnamefont{Kaiser}},
	\bibinfo{author}{\bibfnamefont{S.~A.} \bibnamefont{Werner}},
	\bibinfo{author}{\bibfnamefont{M.}~\bibnamefont{Arif}}, \bibnamefont{and}
	\bibinfo{author}{\bibfnamefont{R.}~\bibnamefont{Clothier}},
	\bibinfo{journal}{Phys. Rev. Lett.} \textbf{\bibinfo{volume}{63}},
	\bibinfo{pages}{380} (\bibinfo{year}{1989}).
	
	\bibitem[{\citenamefont{He and McKellar}(1993)}]{He-Mackellar-PhysRevA}
	\bibinfo{author}{\bibfnamefont{X.-G.} \bibnamefont{He}} \bibnamefont{and}
	\bibinfo{author}{\bibfnamefont{B.~H.~J.} \bibnamefont{McKellar}},
	\bibinfo{journal}{Phys. Rev. A} \textbf{\bibinfo{volume}{47}},
	\bibinfo{pages}{3424} (\bibinfo{year}{1993}).
	
	\bibitem[{\citenamefont{Wilkens}(1994)}]{Wilkens}
	\bibinfo{author}{\bibfnamefont{M.}~\bibnamefont{Wilkens}},
	\bibinfo{journal}{Phys. Rev. Lett.} \textbf{\bibinfo{volume}{72}},
	\bibinfo{pages}{5} (\bibinfo{year}{1994}).
	
	\bibitem[{\citenamefont{Wang and Li}(2007{\natexlab{a}})}]{Wang20072197}
	\bibinfo{author}{\bibfnamefont{J.}~\bibnamefont{Wang}} \bibnamefont{and}
	\bibinfo{author}{\bibfnamefont{K.}~\bibnamefont{Li}}, \bibinfo{journal}{J.
		Phys. A: Math. Theor.} \textbf{\bibinfo{volume}{40}}, \bibinfo{pages}{2197}
	(\bibinfo{year}{2007}{\natexlab{a}}).
	
	\bibitem[{\citenamefont{Wang and Li}(2007{\natexlab{b}})}]{Wang20075}
	\bibinfo{author}{\bibfnamefont{J.-H.} \bibnamefont{Wang}} \bibnamefont{and}
	\bibinfo{author}{\bibfnamefont{K.}~\bibnamefont{Li}},
	\bibinfo{journal}{Chinese Phys. Lett.} \textbf{\bibinfo{volume}{24}},
	\bibinfo{pages}{5} (\bibinfo{year}{2007}{\natexlab{b}}).
	
	\bibitem[{\citenamefont{Dayi}(2009)}]{Dayi2009}
	\bibinfo{author}{\bibfnamefont{O.~F.} \bibnamefont{Dayi}},
	\bibinfo{journal}{EPL} \textbf{\bibinfo{volume}{85}} (\bibinfo{year}{2009}).
	
	\bibitem[{\citenamefont{Anandan}(2000)}]{Anandan-PhysRevLett.85.1354}
	\bibinfo{author}{\bibfnamefont{J.}~\bibnamefont{Anandan}},
	\bibinfo{journal}{Phys. Rev. Lett.} \textbf{\bibinfo{volume}{85}},
	\bibinfo{pages}{1354} (\bibinfo{year}{2000}).
	
	\bibitem[{\citenamefont{Passos et~al.}(2007)\citenamefont{Passos, Ribeiro,
			Furtado, and Nascimento}}]{Pasos-PhysRevA.76.012113}
	\bibinfo{author}{\bibfnamefont{E.}~\bibnamefont{Passos}},
	\bibinfo{author}{\bibfnamefont{L.~R.} \bibnamefont{Ribeiro}},
	\bibinfo{author}{\bibfnamefont{C.}~\bibnamefont{Furtado}}, \bibnamefont{and}
	\bibinfo{author}{\bibfnamefont{J.~R.} \bibnamefont{Nascimento}},
	\bibinfo{journal}{Phys. Rev. A} \textbf{\bibinfo{volume}{76}},
	\bibinfo{pages}{012113} (\bibinfo{year}{2007}).
	
	\bibitem[{\citenamefont{{Spavieri}}(2006)}]{Spavieri-E-P}
	\bibinfo{author}{\bibfnamefont{G.}~\bibnamefont{{Spavieri}}},
	\bibinfo{journal}{Eur. Phys. J. D} \textbf{\bibinfo{volume}{37}},
	\bibinfo{pages}{327} (\bibinfo{year}{2006}).
	
	\bibitem[{\citenamefont{Spavieri and Rodriguez}(2007)}]{spavieri:052113}
	\bibinfo{author}{\bibfnamefont{G.}~\bibnamefont{Spavieri}} \bibnamefont{and}
	\bibinfo{author}{\bibfnamefont{M.}~\bibnamefont{Rodriguez}},
	\bibinfo{journal}{Phys. Rev. A} \textbf{\bibinfo{volume}{75}},
	\bibinfo{eid}{052113} (\bibinfo{year}{2007}).
	
	\bibitem[{\citenamefont{Lepoutre et~al.}(2013)\citenamefont{Lepoutre, Gillot,
			Gauguet, B\"uchner, and Vigu\'e}}]{Lepoutre2013}
	\bibinfo{author}{\bibfnamefont{S.}~\bibnamefont{Lepoutre}},
	\bibinfo{author}{\bibfnamefont{J.}~\bibnamefont{Gillot}},
	\bibinfo{author}{\bibfnamefont{A.}~\bibnamefont{Gauguet}},
	\bibinfo{author}{\bibfnamefont{M.}~\bibnamefont{B\"uchner}},
	\bibnamefont{and} \bibinfo{author}{\bibfnamefont{J.}~\bibnamefont{Vigu\'e}},
	\bibinfo{journal}{Phys. Rev. A} \textbf{\bibinfo{volume}{88}}
	(\bibinfo{year}{2013}).
	
	\bibitem[{\citenamefont{Sangster et~al.}(1993)\citenamefont{Sangster, Hinds,
			Barnett, and Riis}}]{Sangster-PhysRevLett.71.3641}
	\bibinfo{author}{\bibfnamefont{K.}~\bibnamefont{Sangster}},
	\bibinfo{author}{\bibfnamefont{E.~A.} \bibnamefont{Hinds}},
	\bibinfo{author}{\bibfnamefont{S.~M.} \bibnamefont{Barnett}},
	\bibnamefont{and} \bibinfo{author}{\bibfnamefont{E.}~\bibnamefont{Riis}},
	\bibinfo{journal}{Phys. Rev. Lett.} \textbf{\bibinfo{volume}{71}},
	\bibinfo{pages}{3641} (\bibinfo{year}{1993}).
	
	\bibitem[{\citenamefont{Man'ko et~al.}(2007)\citenamefont{Man'ko, Man'ko,
			Marmo, and Vitale}}]{MANKO2007522}
	\bibinfo{author}{\bibfnamefont{O.}~\bibnamefont{Man'ko}},
	\bibinfo{author}{\bibfnamefont{V.}~\bibnamefont{Man'ko}},
	\bibinfo{author}{\bibfnamefont{G.}~\bibnamefont{Marmo}}, \bibnamefont{and}
	\bibinfo{author}{\bibfnamefont{P.}~\bibnamefont{Vitale}},
	\bibinfo{journal}{Phys. Lett. A} \textbf{\bibinfo{volume}{360}},
	\bibinfo{pages}{522 } (\bibinfo{year}{2007}).
	
	\bibitem[{\citenamefont{Dulat and LI}(2006)}]{Bopp-Shift-DULAT-2006}
	\bibinfo{author}{\bibfnamefont{S.}~\bibnamefont{Dulat}} \bibnamefont{and}
	\bibinfo{author}{\bibfnamefont{K.}~\bibnamefont{LI}}, \bibinfo{journal}{Mod.
		Phys. Lett. A} \textbf{\bibinfo{volume}{21}} (\bibinfo{year}{2006}).
	
	\bibitem[{\citenamefont{Chambers}(1960)}]{CHAMBERS-PhysRevLett.5.3}
	\bibinfo{author}{\bibfnamefont{R.~G.} \bibnamefont{Chambers}},
	\bibinfo{journal}{Phys. Rev. Lett.} \textbf{\bibinfo{volume}{5}},
	\bibinfo{pages}{3} (\bibinfo{year}{1960}).
	
	\bibitem[{\citenamefont{Dowling et~al.}(1999)\citenamefont{Dowling, Williams,
			and Franson}}]{DWF-PhysRevLett.83.2486}
	\bibinfo{author}{\bibfnamefont{J.~P.} \bibnamefont{Dowling}},
	\bibinfo{author}{\bibfnamefont{C.~P.} \bibnamefont{Williams}},
	\bibnamefont{and} \bibinfo{author}{\bibfnamefont{J.~D.}
		\bibnamefont{Franson}}, \bibinfo{journal}{Phys. Rev. Lett.}
	\textbf{\bibinfo{volume}{83}}, \bibinfo{pages}{2486} (\bibinfo{year}{1999}).
	
	\bibitem[{\citenamefont{Boulware and Deser}(1989)}]{BD-PhysRevLett.63.2319}
	\bibinfo{author}{\bibfnamefont{D.~G.} \bibnamefont{Boulware}} \bibnamefont{and}
	\bibinfo{author}{\bibfnamefont{S.}~\bibnamefont{Deser}},
	\bibinfo{journal}{Phys. Rev. Lett.} \textbf{\bibinfo{volume}{63}},
	\bibinfo{pages}{2319} (\bibinfo{year}{1989}).
	
	\bibitem[{\citenamefont{Tonomura}(1987)}]{Tonomura1987639}
	\bibinfo{author}{\bibfnamefont{A.}~\bibnamefont{Tonomura}},
	\bibinfo{journal}{Rev. Mod. Phys.} \textbf{\bibinfo{volume}{59}},
	\bibinfo{pages}{639} (\bibinfo{year}{1987}).
	
	\bibitem[{\citenamefont{Neyenhuis et~al.}(2007)\citenamefont{Neyenhuis,
			Christensen, and Durfee}}]{Neyenhuis-Mass-Photon}
	\bibinfo{author}{\bibfnamefont{B.}~\bibnamefont{Neyenhuis}},
	\bibinfo{author}{\bibfnamefont{D.}~\bibnamefont{Christensen}},
	\bibnamefont{and} \bibinfo{author}{\bibfnamefont{D.~S.}
		\bibnamefont{Durfee}}, \bibinfo{journal}{Phys. Rev. Lett.}
	\textbf{\bibinfo{volume}{99}}, \bibinfo{eid}{200401} (\bibinfo{year}{2007}).
	
	\bibitem[{\citenamefont{Rodriguez}(2009)}]{Rodriguez2009373}
	\bibinfo{author}{\bibfnamefont{M.}~\bibnamefont{Rodriguez}},
	\bibinfo{journal}{Rev. Mex. Fis.} \textbf{\bibinfo{volume}{55}},
	\bibinfo{pages}{373} (\bibinfo{year}{2009}).
	
	\bibitem[{\citenamefont{Jin}(2015{\natexlab{a}})}]{1674-1056-24-5-053702}
	\bibinfo{author}{\bibfnamefont{W.}~\bibnamefont{Jin}},
	\bibinfo{journal}{Chinese Phys. B} \textbf{\bibinfo{volume}{24}},
	\bibinfo{pages}{053702} (\bibinfo{year}{2015}{\natexlab{a}}).
	
	\bibitem[{\citenamefont{Gustavson et~al.}(2000)\citenamefont{Gustavson,
			Landragin, and Kasevich}}]{gustavson-2000}
	\bibinfo{author}{\bibfnamefont{T.~L.} \bibnamefont{Gustavson}},
	\bibinfo{author}{\bibfnamefont{A.}~\bibnamefont{Landragin}},
	\bibnamefont{and} \bibinfo{author}{\bibfnamefont{M.~A.}
		\bibnamefont{Kasevich}}, \bibinfo{journal}{Classical and Quantum Gravity}
	\textbf{\bibinfo{volume}{17}}, \bibinfo{pages}{2385} (\bibinfo{year}{2000}).
	
	\bibitem[{\citenamefont{Zhou et~al.}(2016)\citenamefont{Zhou, Zhang, Duan, Ke,
			Shao, and Hu}}]{Zhou-2016}
	\bibinfo{author}{\bibfnamefont{M.-K.} \bibnamefont{Zhou}},
	\bibinfo{author}{\bibfnamefont{K.}~\bibnamefont{Zhang}},
	\bibinfo{author}{\bibfnamefont{X.-C.} \bibnamefont{Duan}},
	\bibinfo{author}{\bibfnamefont{Y.}~\bibnamefont{Ke}},
	\bibinfo{author}{\bibfnamefont{C.-G.} \bibnamefont{Shao}}, \bibnamefont{and}
	\bibinfo{author}{\bibfnamefont{Z.-K.} \bibnamefont{Hu}},
	\bibinfo{journal}{Phys. Rev. A} \textbf{\bibinfo{volume}{93}},
	\bibinfo{pages}{023641} (\bibinfo{year}{2016}).
	
	\bibitem[{\citenamefont{Tkachuk}(2000)}]{Takchuk-PhysRevA.62.052112}
	\bibinfo{author}{\bibfnamefont{V.~M.} \bibnamefont{Tkachuk}},
	\bibinfo{journal}{Phys. Rev. A} \textbf{\bibinfo{volume}{62}},
	\bibinfo{pages}{052112} (\bibinfo{year}{2000}).
	
	\bibitem[{\citenamefont{Moumni et~al.}(2011)\citenamefont{Moumni, BenSlama, and
			Zaim}}]{Moumni2011151}
	\bibinfo{author}{\bibfnamefont{M.}~\bibnamefont{Moumni}},
	\bibinfo{author}{\bibfnamefont{A.}~\bibnamefont{BenSlama}}, \bibnamefont{and}
	\bibinfo{author}{\bibfnamefont{S.}~\bibnamefont{Zaim}}, \bibinfo{journal}{J.
		Geom. Phys.} \textbf{\bibinfo{volume}{61}}, \bibinfo{pages}{151 }
	(\bibinfo{year}{2011}).
	
	\bibitem[{\citenamefont{Lepoutre et~al.}(2012)\citenamefont{Lepoutre, Gauguet,
			Tr\'enec, B\"uchner, and Vigu\'e}}]{Lepoutre2012}
	\bibinfo{author}{\bibfnamefont{S.}~\bibnamefont{Lepoutre}},
	\bibinfo{author}{\bibfnamefont{A.}~\bibnamefont{Gauguet}},
	\bibinfo{author}{\bibfnamefont{G.}~\bibnamefont{Tr\'enec}},
	\bibinfo{author}{\bibfnamefont{M.}~\bibnamefont{B\"uchner}},
	\bibnamefont{and} \bibinfo{author}{\bibfnamefont{J.}~\bibnamefont{Vigu\'e}},
	\bibinfo{journal}{Phys. Rev. Lett.} \textbf{\bibinfo{volume}{109}}
	(\bibinfo{year}{2012}).
	
	\bibitem[{\citenamefont{Jin}(2015{\natexlab{b}})}]{Wang-Jin-2015}
	\bibinfo{author}{\bibfnamefont{W.}~\bibnamefont{Jin}},
	\bibinfo{journal}{Chinese Physics B} \textbf{\bibinfo{volume}{24}},
	\bibinfo{eid}{53702} (pages~\bibinfo{numpages}{0})
	(\bibinfo{year}{2015}{\natexlab{b}}).
	
\end{thebibliography}
\end{document}